\documentclass[twocolumn,showpacs,superscriptaddress,amsmath,amssymb,prl]{revtex4-1}
\usepackage[colorlinks,linkcolor=blue,anchorcolor=blue,citecolor=blue,urlcolor=blue]{hyperref}
\usepackage{graphicx}
\usepackage{epsfig}
\usepackage{dcolumn}
\usepackage{bm}
\usepackage{overpic}
\usepackage{color}
\usepackage{multirow}
\usepackage{subfigure}
\usepackage{ulem}
\usepackage{lineno}
\usepackage{verbatim}
\usepackage{cleveref}

\newcommand\jpsi{J/\psi}
\newcommand\piz{\pi^0}
\newcommand\kpkm{K^+K^-}
\newcommand\az{a_0(980)^0}
\newcommand\fz{f_0(980)}

\begin{document}
\newcommand{\ks}{K_{S}^{0}}
\newcommand{\EP}{e^{+}}
\newcommand{\EM}{e^{-}}
\newcommand{\epm}{e^{\pm}}
\newcommand{\vpho}{\gamma^{\ast}}
\newcommand{\qqbar}{q\bar{q}}

\newcommand{\ee}{e^{+}e^{-}}
\newcommand{\mm}{\mu^{+}\mu^{-}}
\newcommand{\alfs}{\alpha_{s}}
\newcommand{\alfmz}{\alpha(M_{Z}^{2})}
\newcommand{\amu}{a_{\mu}}
\newcommand{\Lam}{\Lambda_{c}}
\newcommand{\lam}{\Lambda_{c}^{+}}
\newcommand{\lambar}{\bar{\Lambda}_{c}^{-}}
\newcommand{\Lambdac}{\Lambda_{c}}
\newcommand{\mbc}{M_{BC}}
\newcommand{\dele}{\Delta E}
\newcommand{\ebm}{E_{\textmd{beam}}}
\newcommand{\ecm}{E_{\textmd{c.m.}}}
\newcommand{\pbm}{p_{\textmd{beam}}}
\newcommand{\MuMu}{\mu\mu}
\newcommand{\mumu}{\mu\mu}
\newcommand{\tata}{\tau^{+}\tau^{-}}
\newcommand{\pipi}{\pi^{+}\pi^{-}}
\newcommand{\gaga}{\gamma\gamma}
\newcommand{\twopho}{\ee+X}
\newcommand{\sqs}{\sqrt{s}}
\newcommand{\sqsp}{\sqrt{s^{\prime}}}
\newcommand{\da}{\Delta\alpha}
\newcommand{\das}{\Delta\alpha(s)}
\newcommand{\dimu}{\ee \ra \mumu}
\newcommand{\dedx}{\textmd{d}E/\textmd{d}x}
\newcommand{\chip}{\chi_{\textmd{Prob}}}
\newcommand{\chiP}{\chi_{p}}
\newcommand{\evz}{V_{z}^{\textmd{evt}}}
\newcommand{\evzloose}{V_{z,\textmd{loose}}^{\textmd{evt}}}
\newcommand{\avz}{V_{z}^{\textmd{ave}}}
\newcommand{\Ngd}{N_{\textmd{good}}}
\newcommand{\Ncru}{N_{\textmd{crude}}}
\newcommand{\pio}{\pi^{0}}
\newcommand{\rpid}{r_{\textmd{PID}}}

\newcommand{\Nhxobs}{N_{h+X}^{\textmd{obs}}}
\newcommand{\Nhobs}{N_{h}^{\textmd{obs}}}
\newcommand{\Npioxobs}{N_{\pi^{0}+X}^{\textmd{obs}}}
\newcommand{\Nksxobs}{N_{\ks+X}^{\textmd{obs}}}
\newcommand{\Npioobs}{N_{\pi^{0}}^{\textmd{obs}}}
\newcommand{\Nksobs}{N_{\ks}^{\textmd{obs}}}
\newcommand{\Nhxtru}{N_{h+X}^{\textmd{tru}}}
\newcommand{\Nhtru}{N_{h}^{\textmd{tru}}}
\newcommand{\Npiotru}{N_{\pi^{0}}^{\textmd{tru}}}
\newcommand{\Nkstru}{N_{\ks}^{\textmd{tru}}}
\newcommand{\Nbarhxobs}{\bar{N}_{h+X}^{\textmd{obs}}}
\newcommand{\Nbarhobs}{\bar{N}_{h}^{\textmd{obs}}}
\newcommand{\Nbarpioobs}{\bar{N}_{\pi^{0}}^{\textmd{obs}}}
\newcommand{\Nbarksobs}{\bar{N}_{\ks}^{\textmd{obs}}}
\newcommand{\Nbarhxtru}{\bar{N}_{h+X}^{\textmd{tru}}}
\newcommand{\Nbarhtru}{\bar{N}_{h}^{\textmd{tru}}}
\newcommand{\Nbarpiotru}{\bar{N}_{\pi^{0}}^{\textmd{tru}}}
\newcommand{\Nbarkstru}{\bar{N}_{\ks}^{\textmd{tru}}}

\newcommand{\Nhadtot}{N_{\textmd{had}}^{\textmd{tot}}}
\newcommand{\Nhadobs}{N_{\textmd{had}}^{\textmd{obs}}}
\newcommand{\Nbarhadobs}{\bar{N}_{\textmd{had}}^{\textmd{obs}}}
\newcommand{\Nhadtru}{N_{\textmd{had}}^{\textmd{tru}}}
\newcommand{\Nbarhadtru}{\bar{N}_{\textmd{had}}^{\textmd{tru}}}
\newcommand{\Nhadphy}{N_{\textmd{had}}}

\newcommand{\cshadobs}{\sigma_{\textmd{had}}^{\textmd{obs}}}
\newcommand{\effhad}{\vap_{\textmd{had}}}
\newcommand{\efftrg}{\vap_{\textmd{trig}}}
\newcommand{\lint}{\mathcal{L}_{\textmd{int}}}
\newcommand{\Nbkg}{N_{\textmd{bkg}}}
\newcommand{\NbkgTot}{N_{\textrm{bkg}}^{\textrm{Tot}}}
\newcommand{\csbkg}{\sigma_{\textmd{bkg}}}
\newcommand{\Nmcsur}{N_{\textmd{MC}}^{\textmd{sur}}}
\newcommand{\Nmcsurori}{N_{\textmd{MC}}^{\textmd{sur,nom.}}}
\newcommand{\Nmcsurwtd}{N_{\textmd{MC}}^{\textmd{sur,wtd.}}}
\newcommand{\Nmcgen}{N_{\textmd{MC}}^{\textmd{gen}}}
\newcommand{\vap}{\varepsilon}
\newcommand{\chisq}{\chi^{2}}
\newcommand{\cshadphy}{\sigma_{\textmd{had}}^{\textmd{phy}}}
\newcommand{\cshadtot}{\sigma_{\textmd{had}}^{\textmd{tot}}}
\newcommand{\cshadborn}{\sigma_{\textmd{had}}^{0}}
\newcommand{\cshadborncon}{\sigma_{\textmd{con}}^{0}}
\newcommand{\cshadbornres}{\sigma_{\textmd{res}}^{0}}
\newcommand{\csdimuborn}{\sigma_{\mu\mu}^{0}}
\newcommand{\rpqcd}{R_{\textmd{pQCD}}}
\newcommand{\Nprod}{N_{\textmd{prod}}}
\newcommand{\Nhadnet}{N_{\textrm{had}}^{\textrm{net}}}
\newcommand{\Delrel}{\Delta_{\textrm{rel}}}

\newcommand{\fourpionchg}{\pipi\pipi}
\newcommand{\fourpionneu}{\pipi\pi^{0}\pi^{0}}
\newcommand{\sixpionchg}{3(\pipi)}
\newcommand{\thrpionneu}{\pipi\pi^{0}}
\newcommand{\twopionchg}{\pipi}

\newcommand{\Nsurnpion}{N_{\textmd{sur}}^{n\pi}}
\newcommand{\Ngennpion}{N_{\textmd{gen}}^{n\pi}}
\newcommand{\Ngentot}{N_{\textmd{gen}}^{\textmd{tot}}}
\newcommand{\effincnpion}{\vap_{n\pi}^{\textmd{inc}}}
\newcommand{\effincnpionp}{\vap_{n\pi}^{\textmd{inc},\prime}}
\newcommand{\effincnonnpion}{\vap_{\textmd{non}-n\pi}^{\textmd{inc}}}
\newcommand{\effexcnpion}{\vap_{n\pi}^{\textmd{exc}}}
\newcommand{\fracnpion}{f_{n\pi}}
\newcommand{\fracnpionp}{f_{n\pi}^{\prime}}
\newcommand{\fracnonnpion}{f_{\textmd{non}-n\pi}}

\newcommand{\Nsurtwopion}{N_{\textmd{sur}}^{2\pi}}
\newcommand{\Ngentwopion}{N_{\textmd{gen}}^{2\pi}}
\newcommand{\effinctwopion}{\vap_{2\pi}^{\textmd{inc}}}
\newcommand{\effinctwopionp}{\vap_{2\pi}^{\textmd{inc},\prime}}
\newcommand{\effincnontwopion}{\vap_{\textmd{non}-2\pi}^{\textmd{inc}}}
\newcommand{\effexctwopion}{\vap_{2\pi}^{\textmd{exc}}}
\newcommand{\fractwopion}{f_{2\pi}}
\newcommand{\fractwopionp}{f_{2\pi}^{\prime}}
\newcommand{\fracnontwopion}{f_{\textmd{non}-2\pi}}

\newcommand{\Nsurthrpion}{N_{\textmd{sur}}^{3\pi}}
\newcommand{\Ngenthrpion}{N_{\textmd{gen}}^{3\pi}}
\newcommand{\effincthrpion}{\vap_{3\pi}^{\textmd{inc}}}
\newcommand{\effincthrpionp}{\vap_{3\pi}^{\textmd{inc},\prime}}
\newcommand{\effincnonthrpion}{\vap_{\textmd{non}-3\pi}^{\textmd{inc}}}
\newcommand{\effexcthrpion}{\vap_{3\pi}^{\textmd{exc}}}
\newcommand{\fracthrpion}{f_{3\pi}}
\newcommand{\fracthrpionp}{f_{3\pi}^{\prime}}
\newcommand{\fracnonthrpion}{f_{\textmd{non}-3\pi}}

\newcommand{\Nsurfourpion}{N_{\textmd{sur}}^{4\pi}}
\newcommand{\Ngenfourpion}{N_{\textmd{gen}}^{4\pi}}
\newcommand{\effincfourpion}{\vap_{4\pi}^{\textmd{inc}}}
\newcommand{\effincfourpionp}{\vap_{4\pi}^{\textmd{inc},\prime}}
\newcommand{\effincnonfourpion}{\vap_{\textmd{non}-4\pi}^{\textmd{inc}}}
\newcommand{\effexcfourpion}{\vap_{4\pi}^{\textmd{exc}}}
\newcommand{\fracfourpion}{f_{4\pi}}
\newcommand{\fracfourpionp}{f_{4\pi}^{\prime}}
\newcommand{\fracnonfourpion}{f_{\textmd{non}-4\pi}}

\newcommand{\Npionprod}{N_{\textmd{prod}}^{4\pi}}
\newcommand{\Ndatasur}{N_{\textmd{data}}^{\textmd{sur}}}
\newcommand{\Nobspion}{N_{\textmd{obs}}^{4\pi}}
\newcommand{\Nhadprod}{N_{\textmd{prod}}^{\textmd{had}}}
\newcommand{\sigmaobs}{\sigma_{\textmd{obs}}}
\newcommand{\effhadp}{\vap_{\textmd{had}}^{\prime}}

\newcommand{\effpion}{\vap_{4\pi}}
\newcommand{\effexcpion}{\vap_{4\pi}^{\textmd{exc}}}
\newcommand{\effincpion}{\vap_{4\pi}^{\textmd{inc}}}
\newcommand{\effincpionI}{\vap_{4\pi}^{\textmd{inc},1}}
\newcommand{\effincpionII}{\vap_{4\pi}^{\textmd{inc},2}}
\newcommand{\effincpionp}{\vap_{4\pi}^{\textmd{inc},\prime}}
\newcommand{\effincremain}{\vap_{\textmd{non}-n\pi}^{\textmd{inc}}}

\newcommand{\fracpion}{f_{4\pi}}
\newcommand{\fracnonpion}{f_{\textmd{non}-4\pi}}
\newcommand{\fracnonpionp}{f_{\textmd{non}-4\pi}^{\prime}}
\newcommand{\fracpionII}{f_{4\pi}^{2}}
\newcommand{\fracpionp}{f_{4\pi}^{\prime}}
\newcommand{\reladiff}{\Delta_{\textmd{rel}}}

\newcommand{\Nsursixpion}{N_{\textmd{sur}}^{6\pi}}
\newcommand{\Ngensixpion}{N_{\textmd{gen}}^{6\pi}}
\newcommand{\effincsixpion}{\vap_{6\pi}^{\textmd{inc}}}
\newcommand{\fracsixpion}{f_{6\pi}}

\newcommand{\etot}{E_{\textmd{tot}}}
\newcommand{\ptot}{p_{\textmd{tot}}}
\newcommand{\plab}{p_{\textmd{Lab}}}
\newcommand{\mpiOI}{M(\pi^{0}_{1})}
\newcommand{\mpiOII}{M(\pi^{0}_{2})}

\newcommand{\widtheeoi}{\varGamma^{\textmd{ee}}_{0,i}}
\newcommand{\widtheeoj}{\varGamma^{\textmd{ee}}_{0,j}}
\newcommand{\widtheeo}{\varGamma^{\textmd{ee}}_{0}}
\newcommand{\widthee}{\varGamma^{\textmd{ee}}}
\newcommand{\widtheeexpi}{\varGamma^{\textmd{ee}}_{\textmd{exp},i}}
\newcommand{\widtheeexp}{\varGamma^{\textmd{ee}}_{\textmd{exp}}}
\newcommand{\widthtoti}{\varGamma^{\textmd{tot}}_{i}}
\newcommand{\widthtot}{\varGamma^{\textmd{tot}}}

\newcommand{\vpqed}{\Pi_{\textmd{QED}}}
\newcommand{\vpqcd}{\Pi_{\textmd{QCD}}}
\newcommand{\vpcon}{\Pi_{\textmd{con}}}
\newcommand{\vpres}{\Pi_{\textmd{res}}}
\newcommand{\vpo}{\Pi_{0}}
\newcommand{\rcon}{R_{\textmd{con}}}
\newcommand{\rres}{R_{\textmd{res}}}
\newcommand{\rexp}{R_{\textmd{exp}}}

\newcommand{\delvert}{\delta_{\textmd{vert}}}
\newcommand{\delvp}{\delta_{\textmd{vac}}}
\newcommand{\delbrem}{\delta_{\gamma}}
\newcommand{\delobs}{\delta_{\textmd{obs}}}
\newcommand{\radiatorsf}{F_{\textmd{SF}}}
\newcommand{\radiatorfd}{F_{\textmd{FD}}}
\newcommand{\DelFD}{\Delta_{\textmd{FD}}}
\newcommand{\DelFDCal}{\Delta_{\textmd{cal}}}
\newcommand{\DelFDcs}{\Delta_{\sigma}}
\newcommand{\DelFDvp}{\Delta_{\textmd{vp}}}

\newcommand{\costh}{\cos\theta}
\newcommand{\costhIprg}{\cos\theta_{\textmd{1prg}}}
\newcommand{\costhIIprg}{\cos\theta_{\textmd{2prg}}}
\newcommand{\costhIIIprg}{\cos\theta_{\textmd{3prg}}}
\newcommand{\costhIVprg}{\cos\theta_{\textmd{4prg}}}
\newcommand{\costhrestprg}{\cos\theta_{\textmd{restprg}}}
\newcommand{\emce}{E^{\textmd{ctrk.}}_{\textmd{emc}}}
\newcommand{\emceIprg}{E^{\textmd{ctrk.}}_{\textmd{emc,1prg}}}
\newcommand{\emceIIprg}{E^{\textmd{ctrk.}}_{\textmd{emc,2prg}}}
\newcommand{\emceIIIprg}{E^{\textmd{ctrk.}}_{\textmd{emc,3prg}}}
\newcommand{\emceIVprg}{E^{\textmd{ctrk.}}_{\textmd{emc,4prg}}}
\newcommand{\emcerestprg}{E^{\textmd{ctrk.}}_{\textmd{emc,restprg}}}
\newcommand{\isocosth}{\cos\theta_{\textmd{iso}}}
\newcommand{\isocosthIprg}{\cos\theta_{\textmd{iso,1prg}}}
\newcommand{\isocosthIIprg}{\cos\theta_{\textmd{iso,2prg}}}
\newcommand{\isocosthIIIprg}{\cos\theta_{\textmd{iso,3prg}}}
\newcommand{\isocosthIVprg}{\cos\theta_{\textmd{iso,4prg}}}
\newcommand{\isocosthrestprg}{\cos\theta_{\textmd{iso,restprg}}}
\newcommand{\eop}{E/P}
\newcommand{\eopIprg}{E/P_{\textmd{1prg}}}
\newcommand{\eopIIprg}{E/P_{\textmd{2prg}}}
\newcommand{\eopIIIprg}{E/P_{\textmd{3prg}}}
\newcommand{\eopIVprg}{E/P_{\textmd{4prg}}}
\newcommand{\eoprestprg}{E/P_{\textmd{restprg}}}
\newcommand{\nisogam}{N_{\textmd{isogam}}}
\newcommand{\nisogamIprg}{N_{\textmd{isogam,1prg}}}
\newcommand{\nisogamIIprg}{N_{\textmd{isogam,2prg}}}
\newcommand{\nisogamIIIprg}{N_{\textmd{isogam,3prg}}}
\newcommand{\nisogamIVprg}{N_{\textmd{isogam,4prg}}}
\newcommand{\nisogamrestprg}{N_{\textmd{isogam,restprg}}}
\newcommand{\ptrk}{p_{\textmd{ctrk}}}
\newcommand{\pIprg}{p^{\textmd{ctrk}}_{\textmd{1prg}}}
\newcommand{\pIIprg}{p^{\textmd{ctrk}}_{\textmd{2prg}}}
\newcommand{\pIIIprg}{p^{\textmd{ctrk}}_{\textmd{3prg}}}
\newcommand{\pIVprg}{p^{\textmd{ctrk}}_{\textmd{4prg}}}
\newcommand{\prestprg}{p^{\textmd{ctrk}}_{\textmd{restprg}}}
\newcommand{\tote}{E_{\textmd{vis.}}}
\newcommand{\totevte}{E_{\textmd{tot.}}}
\newcommand{\totevteIprg}{E_{\textmd{tot.}}^{\textmd{1prg}}}
\newcommand{\balanceIprg}{\textmd{Balance}}
\newcommand{\ngamma}{N_{\gamma}}
\newcommand{\ngammaIprg}{N_{\gamma,\textmd{1prg}}}
\newcommand{\ngammaIIprg}{N_{\gamma,\textmd{2prg}}}
\newcommand{\ngammaIIIprg}{N_{\gamma,\textmd{3prg}}}
\newcommand{\ngammaIVprg}{N_{\gamma,\textmd{4prg}}}
\newcommand{\ngammarestprg}{N_{\gamma,\textmd{restprg}}}
\newcommand{\ngoodwt}{N_{\textmd{good}}^{\textmd{Wt}}}
\newcommand{\ngood}{N_{\textmd{good}}}
\newcommand{\npiO}{N_{\pi^{0}}}
\newcommand{\npiOIprg}{N_{\pi^{0}}^{\textmd{1prg}}}
\newcommand{\npiOIIprg}{N_{\pi^{0}}^{\textmd{2prg}}}
\newcommand{\npp}{N_{p}}
\newcommand{\npm}{N_{\bar{p}}}
\newcommand{\nkp}{N_{K^{+}}}
\newcommand{\nkm}{N_{K^{-}}}
\newcommand{\npip}{N_{\pi^{+}}}
\newcommand{\npim}{N_{\pi^{-}}}
\newcommand{\ppp}{P(p^{+})}
\newcommand{\ppm}{P(\bar{p}^{-})}
\newcommand{\pkp}{p(K^{+})}
\newcommand{\pkm}{p(K^{-})}
\newcommand{\ppip}{P(\pi^{+})}
\newcommand{\ppim}{P(\pi^{-})}
\newcommand{\ppiO}{P(\pi^{0})}
\newcommand{\mpiO}{M(\pi^{0})}
\newcommand{\mks}{M(K^{0}_{s})}
\newcommand{\pks}{p_{K^{0}_{s}}}
\newcommand{\mphi}{M(\phi)}
\newcommand{\pphi}{p_{\phi}}
\newcommand{\mIIgam}{M(\gamma\gamma)}
\newcommand{\mIIgamIprg}{M(\gamma\gamma)^{\textmd{1prg}}}
\newcommand{\pIIgam}{p_{\gamma\gamma}}
\newcommand{\mlambda}{M(\Lambda)}
\newcommand{\plambda}{p_{\Lambda}}
\newcommand{\mdO}{M(D^{0})}
\newcommand{\pdO}{p_{D^{0}}}
\newcommand{\mdstarO}{M(D^{\ast 0})}
\newcommand{\pdstarO}{p_{D^{\ast 0}}}
\newcommand{\mdp}{M(D^{\pm})}
\newcommand{\pdp}{p_{D^{\pm}}}
\newcommand{\mdstarp}{M(D^{\ast\pm})}
\newcommand{\pdstarp}{p_{D^{\ast\pm}}}
\newcommand{\mds}{M(D_{s}^{\pm})}
\newcommand{\pds}{p_{D_{s}^{\pm}}}
\newcommand{\mdstars}{M(D_{s}^{\ast\pm})}
\newcommand{\pdstars}{p_{D_{s}^{\ast\pm}}}
\newcommand{\Vr}{V_{r}}
\newcommand{\Vz}{V_{z}}

\newcommand{\gev}{\mathrm{GeV}}
\newcommand{\mev}{\mathrm{MeV}}
\newcommand{\mevcc}{\mathrm{MeV}/c^{2}}
\newcommand{\gevc}{\mathrm{GeV}/c}
\newcommand{\gevcc}{\mathrm{GeV}/c^2}

\newcommand{\nchg}{N_{\textmd{chg}}}
\newcommand{\eff}{\vap}

\newcommand{\critecm}{1.780}

\newcommand{\ENERGYAT}{4575.5}
\newcommand{\ENERGYBT}{4575.5}
\newcommand{\ENERGYCT}{4575.5}
\newcommand{\ENERGYDT}{4575.5}
\newcommand{\ksdecay}{\ks\ra\pi^{+}\pi^{-}}
\newcommand{\phidecay}{\phi\ra K^{+}K^{-}}
\newcommand{\piOdecay}{\pi^{0}\ra\gamma\gamma}
\newcommand{\Lambdadecay}{\Lambda\ra p\pi^{-}}
\newcommand{\DOdecay}{D^{0}\ra K^{-}\pi^{+}}
\newcommand{\DStarOdecay}{D^{\ast0}\ra D^{0}\pi^{0}}
\newcommand{\Dpdecay}{D^{+}\ra K^{+}\pi^{+}\pi^{-}}
\newcommand{\DStarpdecay}{D^{\ast+}\ra D^{0}\pi^{+}}
\newcommand{\Dsdecay}{D^{+}_{s}\ra K^{+}K^{-}\pi^{+}}
\newcommand{\DStarsdecay}{D^{\ast+}_{s}\ra D^{+}_{s}\gamma}


\title{\boldmath \textbf{Study of the decay $\jpsi \to \phi \piz\eta$} }
\author{
\begin{small}
\begin{center}
M.~Ablikim$^{1}$, M.~N.~Achasov$^{4,b}$, P.~Adlarson$^{75}$, X.~C.~Ai$^{80}$, R.~Aliberti$^{35}$, A.~Amoroso$^{74A,74C}$, M.~R.~An$^{39}$, Q.~An$^{71,58}$, Y.~Bai$^{57}$, O.~Bakina$^{36}$
% ...
\\
\vspace{0.2cm}
(BESIII Collaboration)\\
\vspace{0.2cm} {\it
$^{1}$ Institute of High Energy Physics, Beijing 100049, People's Republic of China\\
% ...
$^{4}$ Budker Institute of Nuclear Physics SB RAS (BINP), Novosibirsk 630090, Russia\\
% ...
$^{35}$ Johannes Gutenberg University of Mainz, Johann-Joachim-Becher-Weg 45, D-55099 Mainz, Germany\\
$^{36}$ Joint Institute for Nuclear Research, 141980 Dubna, Moscow region, Russia\\
% ...
$^{39}$ Liaoning Normal University, Dalian 116029, People's Republic of China\\
% ...
$^{57}$ Southeast University, Nanjing 211100, People's Republic of China\\
$^{58}$ State Key Laboratory of Particle Detection and Electronics, Beijing 100049, Hefei 230026, People's Republic of China\\
% ...
$^{71}$ University of Science and Technology of China, Hefei 230026, People's Republic of China\\
% ...
$^{74}$ University of Turin and INFN, (A)University of Turin, I-10125, Turin, Italy; (B)University of Eastern Piedmont, I-15121, Alessandria, Italy; (C)INFN, I-10125, Turin, Italy\\
$^{75}$ Uppsala University, Box 516, SE-75120 Uppsala, Sweden\\
% ...
$^{80}$ Zhengzhou University, Zhengzhou 450001, People's Republic of China\\
\vspace{0.2cm}
% ...
$^{b}$ Also at the Novosibirsk State University, Novosibirsk, 630090, Russia\\
% ...
}
\end{center}
\vspace{0.4cm}
\vspace{0.4cm}
\end{small}
}

\date{\today}

\begin{abstract}

Based on $(10.09 \pm 0.04) \times 10^9$ $\jpsi$ events collected with the BESIII detector operating at the BEPCII collider, a partial wave analysis of the decay $\jpsi \to \phi \piz \eta$ is performed.
We observe for the first time two new structures  on the $\phi\eta$ invariant mass distribution, with statistical significances of $24.0\sigma$ and $16.9\sigma$;
the first with $J^{\rm PC}$ = $1^{+-}$, mass M = (1911 $\pm$ 6 (stat.) $\pm$ 14 (sys.))~$\mevcc$, and width $\Gamma = $ (149 $\pm$ 12 (stat.) $\pm$ 23 (sys.))~$\mev$, the second with $J^{\rm PC}$ = $1^{--}$, mass M = (1996 $\pm$ 11 (stat.) $\pm$ 30 (sys.))~$\mevcc$, and width $\Gamma$ = (148 $\pm$ 16 (stat.) $\pm$ 66 (sys.))~$\mev$.
These measurements provide important input for the strangeonium spectrum.
In addition, the $\fz-\az$ mixing signal in $\jpsi \to \phi\fz \to \phi\az$ and the corresponding  electromagnetic decay $\jpsi \to \phi\az$ are measured with improved precision, providing crucial information to understand the nature of $\az$ and $\fz$.

\end{abstract}

\maketitle



Studies of the strangeonium ($s\bar{s}$) spectrum and of exotic states containing $s$ and $\bar{s}$ quarks offer a unique platform to deepen our understanding of Quantum Chromodynamics (QCD) in the nonperturbative regime~\cite{Godfrey:1985xj, ParticleDataGroup:2020ssz}.
At present, our knowledge of the $s\bar{s}$ spectrum is incomplete and only a small fraction of the predicted $s\bar{s}$ states is confirmed in the mass region below 2.2~$\gevcc$~\cite{ParticleDataGroup:2020ssz,Barnes:2002mu,Li:2020xzs}. 
Although the existence of many of these states has been studied by experiments, their nature is still controversial.
The lightest axial-vector state $h_1(1415)$ was observed by both LASS~\cite{Aston:1987ak} and Crystal Barrel~\cite{CrystalBarrel:1997kda} collaborations about thirty years ago, and its parameters were recently measured with improved precision by BESIII~\cite{BESIII:2015vfb,BESIII:2018ede}. 
The possible interpretations of its internal nature include a mixture between $SU(3)$-singlet $1^{1}P_{1}$ and $SU(3)$-octet  $1^{3}P_{1}$  mesons~\cite{Li:2005eq} or a $q\bar{q}$ state as predicted by the relativistic quark model~\cite{Godfrey:1985xj}.
Its excitations, e.g., $h_1(2P)$ and $h_1(3P)$, have been theoretically predicted~\cite{Barnes:2002mu}. Their measurements, however, are still limited by the current experimental sensitivity.
The search for these $s\bar{s}$ excited states and their systematical investigation are crucial to improve our understanding of the $h_1$ family~\cite{Chen:2015iqa}.
Since gluons and light quarks are produced abundantly in the decay of charmonium states, they offer an excellent opportunity for $s\bar{s}$ spectroscopy.
Different approaches to search for new $s\bar{s}$ states in charmonium decays were proposed~\cite{Barnes:2002mu}. 

The light scalar mesons $\az$ and $\fz$ are considered as candidates for exotic states. The mixing mechanism of the  $\az - \fz$ system provides an important access to understand their nature~\cite{Achasov:1979xc,Close:2000ah,Close:2001ay,Hanhart:2007bd,Wu:2007jh,Wu:2008hx}.
By using $1.31 \times 10^9$ $\jpsi$ and $4.48 \times 10^{8}$ $\psi(3686)$ events, BESIII has discovered $\az - \fz$ mixing signals with the statistical significance of 7.4$\sigma$ and 5.5$\sigma$ in the decays $\jpsi \to \phi\piz\eta$ and $\chi_{c1} \to \pi^{+}\pi^{-}\piz$~\cite{BESIII:2018ozj}, respectively, where the destructive and constructive solutions of the mixing intensities are presented, too.
A more precise measurement of the mixing intensity will improve our understanding of the nature of $\az$ and $\fz$.

In this Letter, a partial wave analysis (PWA) on the isospin violating decay $\jpsi \to \phi\piz \eta$  is performed  based on $(10.09 \pm 0.04) \times 10^9$ $\jpsi$ events~\cite{BESIII:2021cxx} accumulated with the BESIII detector operating at the BEPCII collider. Two new structures are observed on the $\phi\eta$ invariant mass distribution, and their parameters are hereby determined. In addition, the mixing of the $\fz\to\az$ signal is confirmed, and the corresponding mixing intensity is measured with improved precision.

The BESIII detector records symmetric $e^+e^-$ collisions  provided by the BEPCII storage ring~\cite{Yu:IPAC2016-TUYA01}
in the center-of-mass energy range from 2.0 to 4.95~GeV, with a peak luminosity of $1 \times 10^{33}\;\text{cm}^{-2}\text{s}^{-1}$ 
achieved at $\sqrt{s} = 3.77\;\text{GeV}$.
The detailed description of the BESIII detector can be found in Ref.~\cite{Ablikim:2009aa}.
The simulated Monte Carlo (MC)  samples based on the {\sc geant4}~\cite{GEANT4:2002zbu} package, incorporating the geometric description of the BESIII detector~\cite{detvis} and the detector response, are used to determine the detection efficiencies and to estimate the background. 
In the simulation,
the beam energy spread and the initial state radiation in the $e^+e^-$ annihilation are incorporated with the generator {\sc kkmc}~\cite{Jadach:2000ir,Jadach:1999vf}. An inclusive MC sample,  including both the production of
the $\jpsi$ resonance and the continuum processes, is generated with {\sc kkmc}~\cite{Jadach:2000ir,Jadach:1999vf} for the same number of events as in real data. The known decay modes are incorporated with {\sc evtgen}~\cite{Lange:2001uf,Ping:2008zz} taking the branching fractions from the Particle Data Group (PDG)~\cite{ParticleDataGroup:2020ssz}, while the remaining unknown charmonium decays are modelled with {\sc lundcharm}~\cite{Chen:2000tv,Yang:2014vra}. Final state radiation from charged particles in the final state is incorporated with {\sc photos}~\cite{Barberio:1993qi}. Signal MC samples for the process $\jpsi \to \phi\piz\eta$ with the subsequent decays $\phi \to \kpkm$, $\piz \to \gamma\gamma$, and $\eta \to \gamma\gamma$ are generated uniformly in phase space. 

Candidate events for the decay $\jpsi \to \phi \piz \eta$ are selected using the criteria described in Ref.~\cite{BESIII:2018ozj}, where the $\phi$ signal is reconstructed by the $K^{+}K^{-}$ decay and the $\piz$ and $\eta$ mesons are reconstructed by the $\gamma\gamma$ decay mode.
A four-constraint (4C) kinematic fit enforcing energy-momentum conservation is carried out under the hypothesis of $\jpsi\to K^{+}K^{-}\gamma\gamma\gamma\gamma$. For the events with more
than four photons, the combination with the smallest $\chi_{4\rm C}^2$ is kept, and $\chi^2_{4\rm C} <45$ is additionally required. The $\piz$ and $\eta$ candidates are reconstructed by minimizing $\chi^2_{\piz\eta} =(M_{\gamma_1\gamma_2}-m_{\piz})^2/\sigma_{\piz}^2+(M_{\gamma_3\gamma_4}-m_{\eta})^2/\sigma_{\eta}^2$, where $m_{\piz}$ and $m_{\eta}$ are the nominal mass values of $\piz$ and $\eta$ quoted from the PDG~\cite{ParticleDataGroup:2020ssz}, $M_{\gamma_1\gamma_2}$ and $M_{\gamma_3\gamma_4}$ are the invariant masses of the $\gamma_1\gamma_2$ and $\gamma_3\gamma_4$ combinations, and $\sigma_{\piz}$ and $\sigma_{\eta}$ are their corresponding resolutions estimated by the MC sample $\jpsi \to \phi\piz\eta$ to be 4.4~$\mevcc$ and 7.3~$\mevcc$, respectively. The $\piz$ and $\eta$ candidates are selected by requiring $|M_{\gamma_1\gamma_2}-m_{\piz}| < 15~\mevcc$ and  $|M_{\gamma_3\gamma_4}-m_{\eta}| < 25~\mevcc$, respectively. To reject the background from $\jpsi \to K^{+}K^{-}\piz\piz$ and $\jpsi \to K^{+}K^{-}\eta\eta$, two $\chi^2$ functions analogous to $\chi^2_{\piz\eta}$ are defined for the $\piz\piz$ and $\eta\eta$ pairs, requiring that $\chi^2_{\piz\piz}> 90$ and  $\chi^2_{\eta\eta}> 8$. The $\chi^2$ requirements are optimized to maximize
the figure of merit  $S/\sqrt{S+B}$, where $S$ is the signal and $B$ is the background.
The $\phi$ candidates are selected by requiring the $\kpkm$ invariant mass being in the range $|M_{\kpkm}-m_{\phi}| < 10~\mevcc$, where $m_{\phi}$ is the $\phi$ nominal mass.
 To improve the kinematic resolution, a six-constraint (6C) kinematic fit is performed, using the $\eta$ and $\piz$ masses taken from PDG~\cite{ParticleDataGroup:2020ssz} as well as energy and momentum conservation as constraints. 
 The resulting kinematic variables are used in the further analysis. 

\begin{figure}[!htbp]
 \centering
 \includegraphics[width=8.0cm]{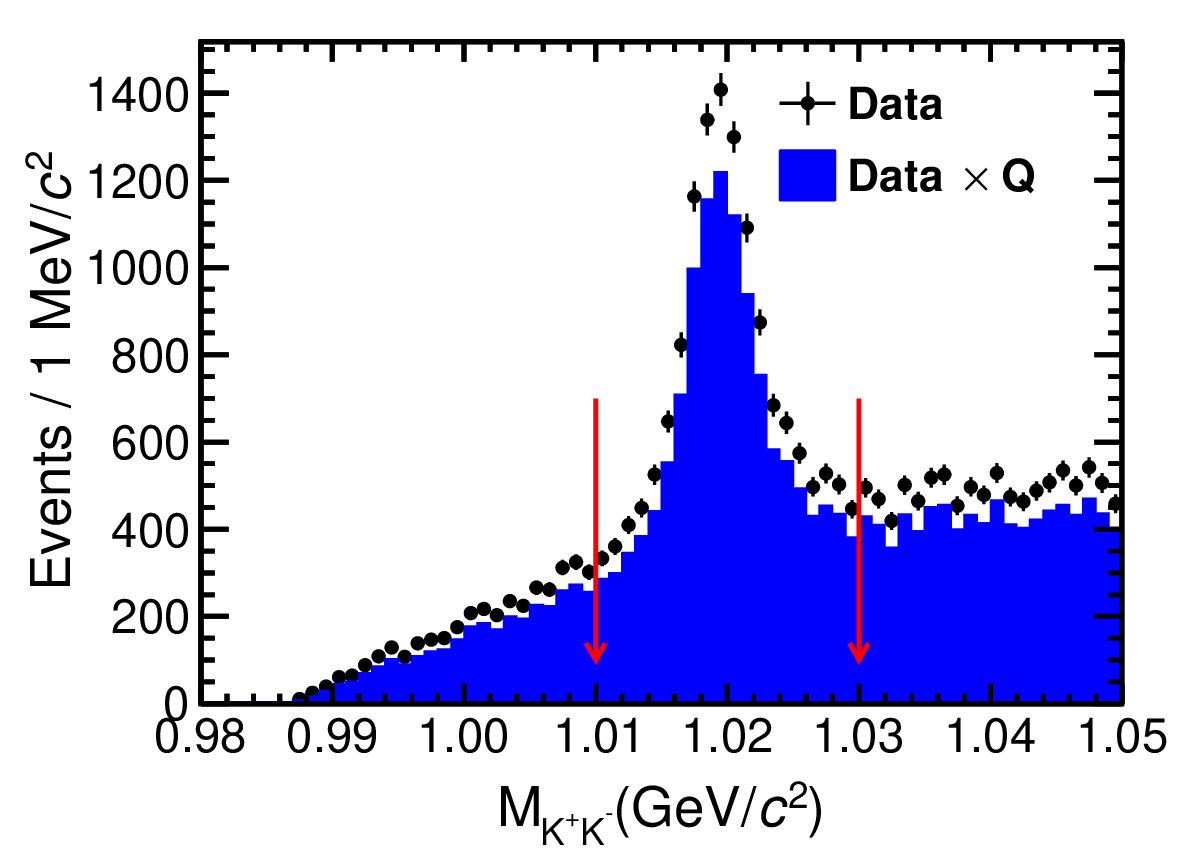}
    \vspace*{-0.3cm}
 \caption{ Distribution of the $\kpkm$ invariant mass. The black dots are data, the blue-shaded area shows the $Q$-weighted distribution. }
 \label{mkk:1}
\end{figure}

After applying all the above selection criteria, the distribution of the $\kpkm$ invariant mass ($M_{\kpkm}$) of the remaining events is shown in Fig.~\ref{mkk:1}, where the $\phi$ signal is observed clearly together with a considerable non-$\phi$ background.
Detailed studies of MC simulation indicate that the background includes also contributions with no $\piz$ and no $\eta$ mesons in the final state. Therefore, a sophisticated $Q$-factor method~\cite{Williams:2008sh} is applied to subtract the corresponding background.
For each event, a weight factor $Q_i$ is assigned, that is determined by analyzing the signal-to-background ratio by fitting the two-dimensional distribution $M_{\gamma_1\gamma_2}$ versus $M_{\gamma_3\gamma_4}$ in a very small cell of the available phase space around each event.
The corresponding $Q$-weighted $M_{\kpkm}$ distribution is shown as the shaded plot in Fig.~\ref{mkk:1}. From the $M_{\kpkm}$ distribution we obtain 14585 events within the $\phi$ signal region ($1.01<M_{\kpkm}<1.03~\gevcc$) without weighting, and 12561 events after the $Q$-weighting. The weighted $M_{\gamma\gamma}$ distributions are shown in the supplementary materials~\cite{suppl}.
To validate the $Q$-factor method, the number of non-$\piz$ and non-$\eta$ background events is also estimated by analyzing events in the $\piz$ and $\eta$ sideband regions, and an agreement of $\sim0.6\%$ on the number of background events in the signal region between the two methods is achieved.

\begin{figure*}[htbp]
 \centering
 \begin{overpic}[width=0.32\textwidth]{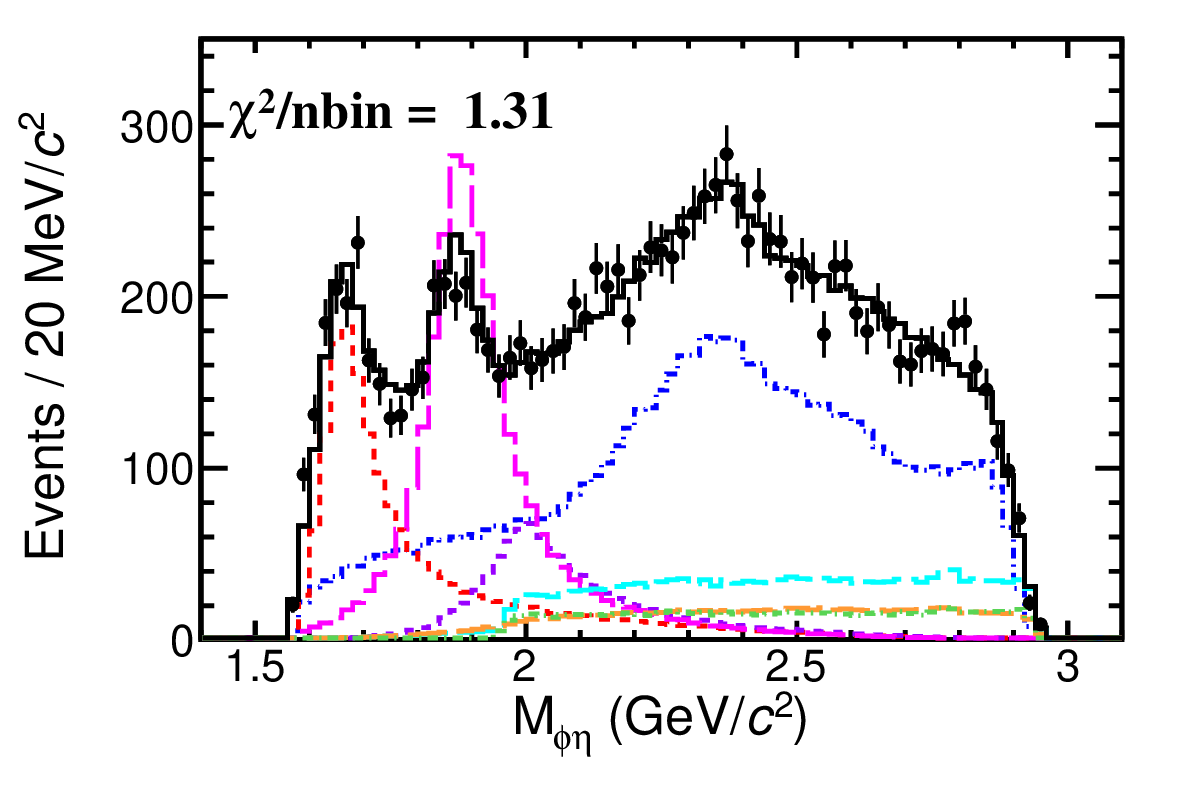}
 \put(65,55){ (a)}
 \end{overpic}
 \begin{overpic}[width=0.32\textwidth]{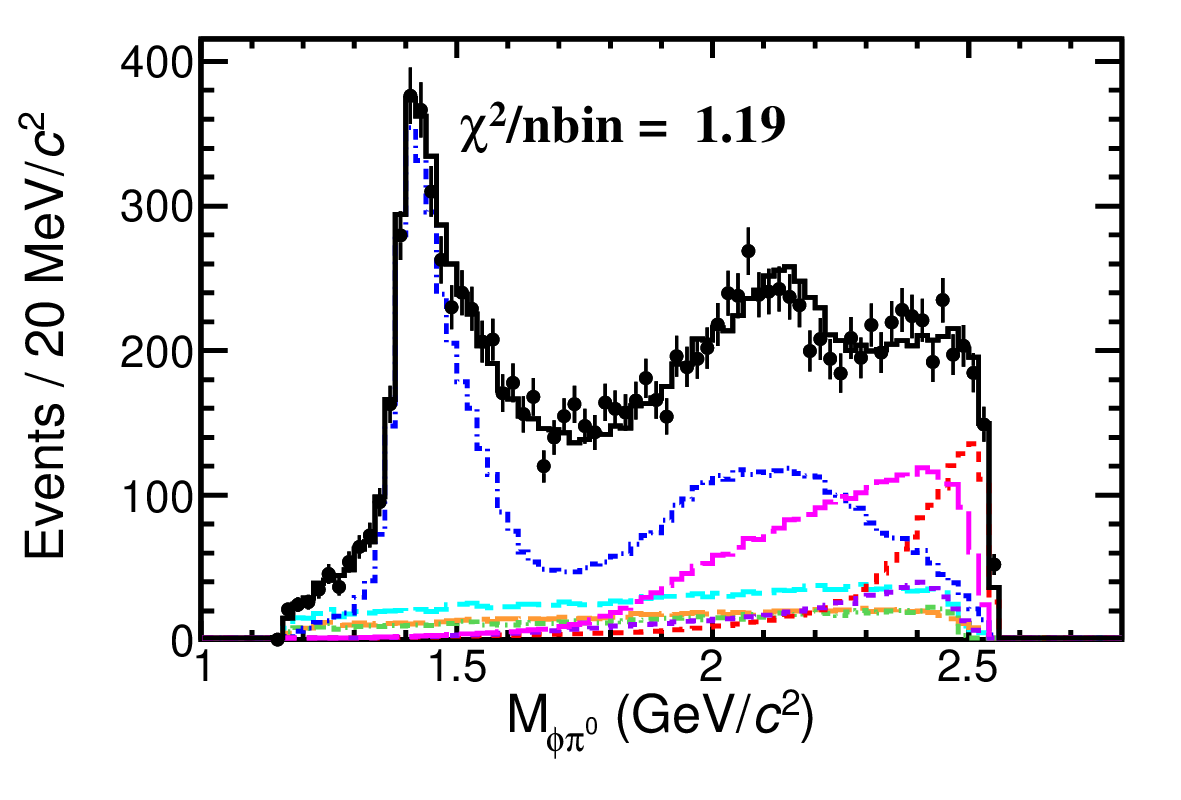}
 \put(70,55){ (b)}
 \end{overpic}
 \begin{overpic}[width=0.32\textwidth]{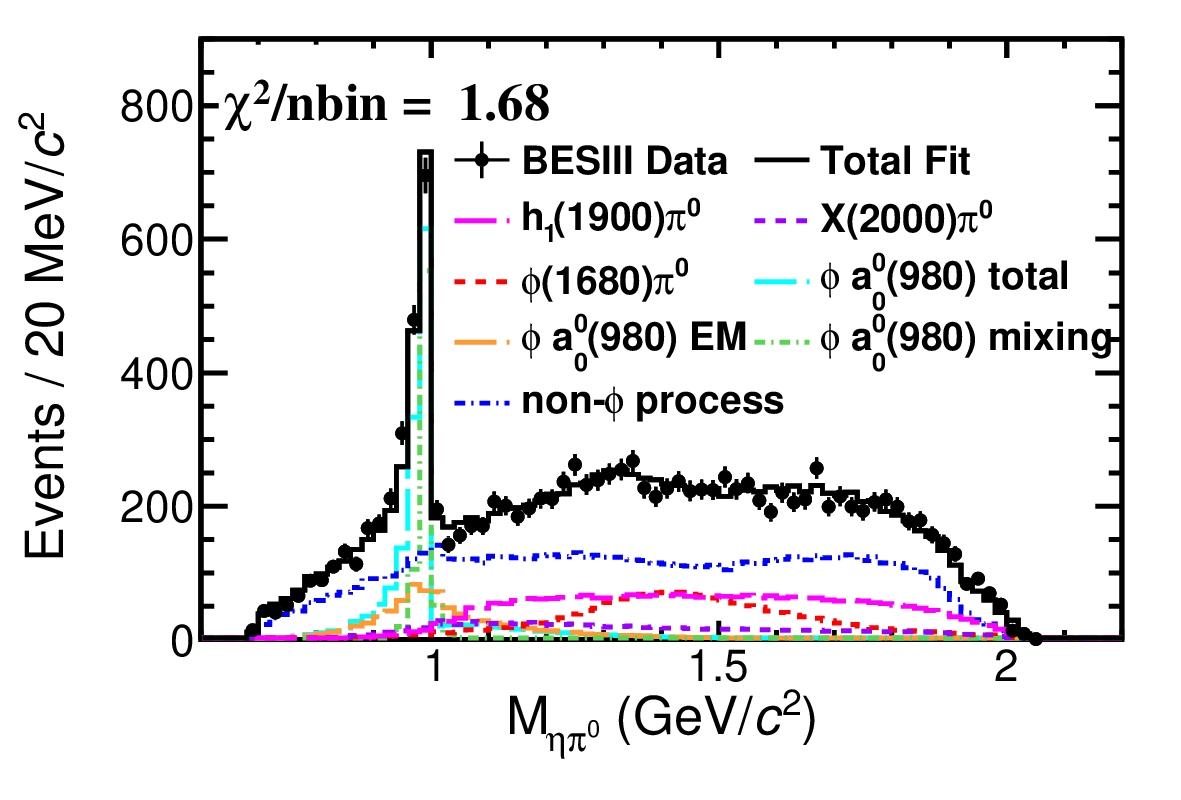}
 \put(83,55){ (c)}
 \end{overpic}

 \begin{overpic}[width=0.32\textwidth]{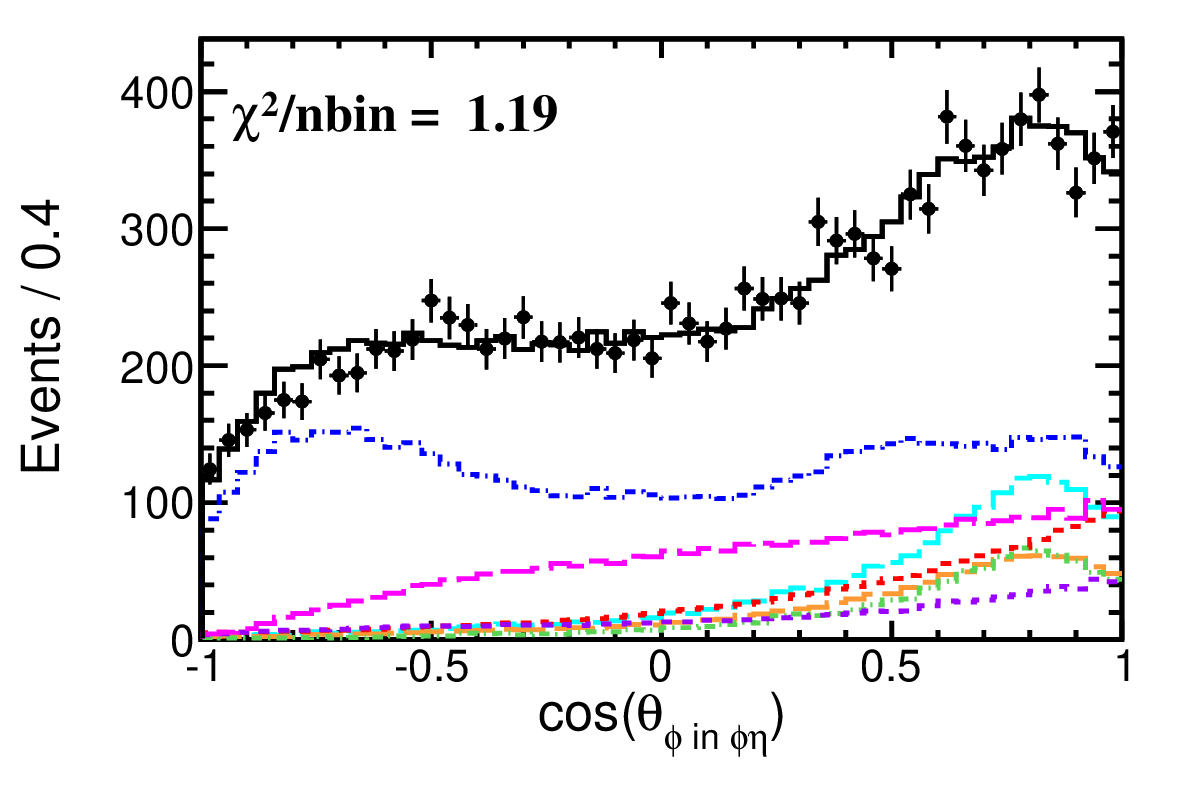}
 \put(65,55){ (d)}
 \end{overpic}
 \begin{overpic}[width=0.32\textwidth]{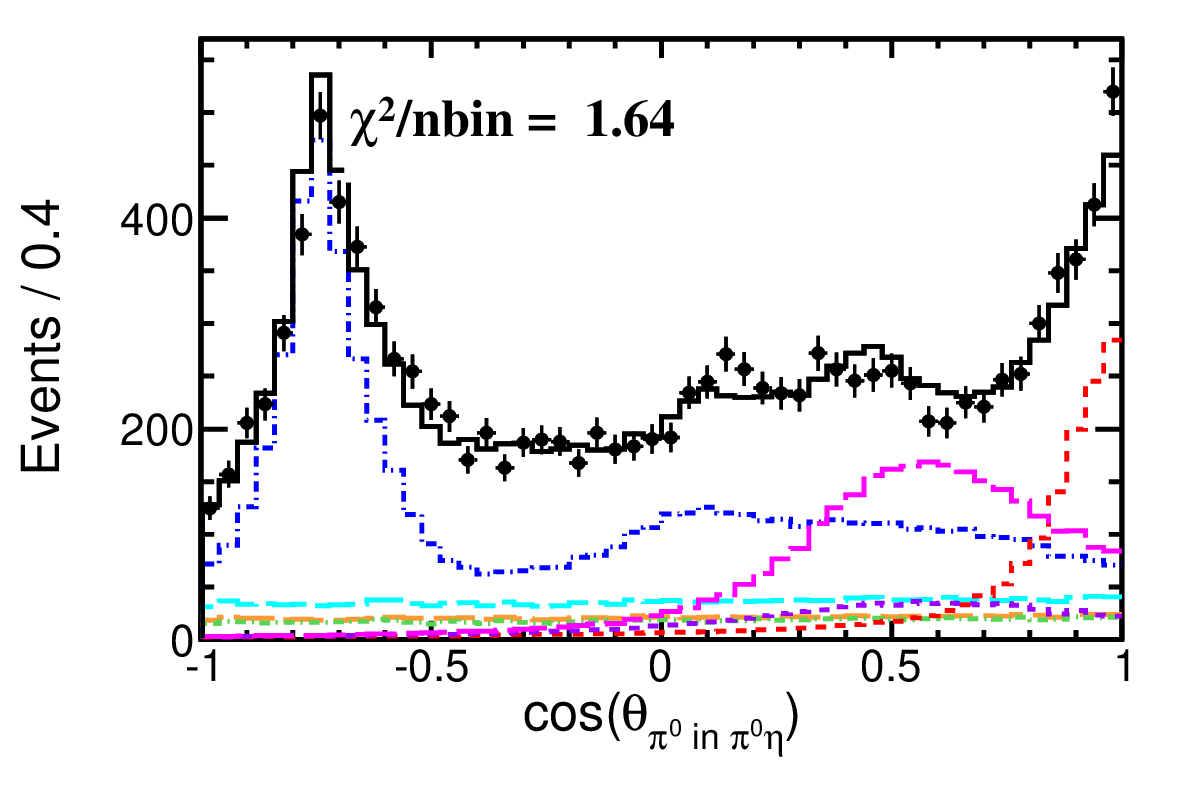}
 \put(70,55){ (e)}
 \end{overpic}
 \begin{overpic}[width=0.32\textwidth]{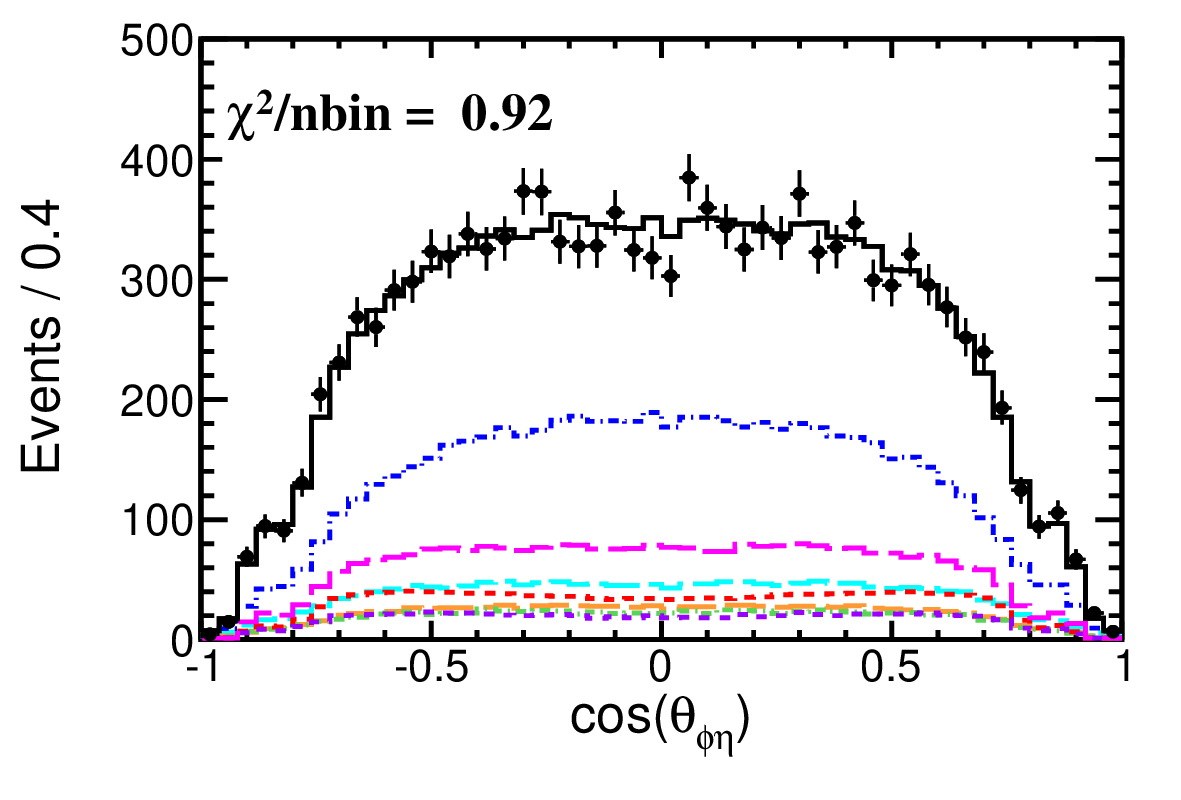}
 \put(83,55){ (f)}
 \end{overpic}
 \caption{
 Invariant mass distributions of (a) $\phi\eta$, (b) $\phi\piz$, and (c) $\eta\piz$, and angular distributions of (d) cos$\theta_\phi$ in the $\phi\eta$ helicity frame, (e) cos$\theta_{\piz}$ in the $\piz\eta$ helicity frame, (f) cos$\theta_{\phi\eta}$ in the center-of-mass rest frame.
 The black dots are the background-subtracted data, the black continuous lines are the  PWA fit projections, the colored non-continuous lines show the components of the fit model.
 }
 \label{PWAfiguure2125:2}
\end{figure*}

Figure~\ref{PWAfiguure2125:2} shows the distributions of  $\phi\eta$, $\phi\piz$, and $\eta\piz$ invariant mass ($M_{\phi\eta}$, $M_{\phi\piz}$, and $M_{\eta\piz}$) and the corresponding angular distributions, after the $Q$-weighting, for the candidate events within the $\phi$ signal region.
 Besides the well established $\az-\fz$ mixing signal in the $M_{\eta\piz}$ distribution~\cite{BESIII:2018ozj}, two structures with  $M_{\phi\eta}$ lower than 2~$\gevcc$ and a prominent peak at $M_{\phi\piz}$ around 1.4~$\gevcc$ are observed clearly.
 A similar peak around 1.4~$\gevcc$ in the $M_{\phi\piz}$ distribution is observed for the events in the $\phi$ sideband region, defined as $1.05<M_{\kpkm}<1.07~\gevcc$, shown in the  supplementary materials~\cite{suppl}.
This indicates that this structure is not from the $\phi\piz$ decay mode, but from the $\kpkm\piz$ channel.

To identify the properties of the new observed structures and precisely measure the intensity of the $\az-\fz$ mixing signal, a PWA based on the GPUPWA framework~\cite{Berger:2010zza} is carried out. 
Detailed studies on the candidate events in the $\phi$ sideband region and in the MC simulation indicate that the components of the non-$\phi$ background are very complicated due to the $K^{*(\prime)}$ contributions.
Therefore, to better model the non-$\phi$ background,  a similar PWA is performed on the candidate events with $M_{\kpkm}$ in the range between 1.07 and 1.5~$\gevcc$, including the different  $K^{*(\prime)}$  components.
The  supplementary materials~\cite{suppl}  show the invariant mass and angular distributions of data used in the above PWA and the MC simulation projection based on the PWA results,
as well as the comparison of the invariant mass and angular distributions for the events in the $\phi$ sideband region between the data and the MC projections. The MC projections describe the data very well.

In the PWA, the two-body decay amplitudes in the sequential decays are constructed using the covariant tensor amplitudes as in Ref.~\cite{Zou:2002ar}. 
Except for the $\az$ state, the $\fz \to \az$ mixing signal and the $\phi(1680)$ state, the intermediate states are parameterized with relativistic Breit-Wigner (BW) functions using the quoted masses and widths from the PDG.
The $\fz$ and $\az$ states are described by the Flatt\'{e} formula with fixed parameters as in Refs.~\cite{BES:2004twe,BESIII:2016tqo}. 
The $\fz \to \az$ mixing signal is parameterized as in Ref.~\cite{BESIII:2018ozj}. 
The $\phi(1680)$ state, whose mass is close to the $\phi\eta$ and $K^*K$ mass thresholds, is parameterized with a three-channel-coupled formula ($\phi\eta$, $K^{*}K$, and $\phi\pi\pi$), in which the coupled parameters are fixed according to Ref.~\cite{BaBar:2011btv}, and its mass and width are free parameters in the fit. 
The resolution effect on the narrow signals, i.e. the $\phi(1020)$ meson and the $\fz \to \az$ mixing signal, is taken into account by convolving the propagator with two independent Gaussian functions determined by the signal MC sample.  
The total decay amplitude for the $i$-th candidate event $M(\xi_i)$ with kinematic variables $\xi_i$ is the sum of the amplitudes of all possible decay processes $A_j(\xi_i)$ together with the amplitude for the non-$\phi$ background $B(\xi_i)$, $M(\xi_i) = \sum_j A_j(\xi_i)+ B(\xi_i)$, where $A_j(\xi_i)$ and $B(\xi_i)$ are the amplitudes for the $\jpsi$ decay process including intermediate resonances.
The complex coefficients of the amplitudes (relative magnitudes and phases) and the resonance parameters (mass and width) in $A_j(\xi_i)$ are obtained by minimizing
\begin{equation}
      S(\xi)= -\ln\mathcal{L} =
      -\alpha\sum_{i}^{N}Q_i\ln\left(\frac{|M(\xi_i)|^2\varepsilon(\xi_{i})}{\sigma'}\right)
 \label{PWAf:6}
 \end{equation}
with an unbinned maximum likelihood fit using MINUIT~\cite{James:1975dr}.
Here, $\varepsilon(\xi_{i})$ is the corresponding detection efficiency, 
$\sigma'=\int |M(\xi)|^2\varepsilon(\xi)d\xi$ is the normalization integral and $Q_i$ is the $Q$-factor of each event. The global factor $\alpha=\sum_{i=1}^{N}Q_i/\sum_{i=1}^{N}Q_i^{2}$ is introduced to take into account the correction on the statistical uncertainty for the weighted likelihood function~\cite{Qf1,Qf2}. 
The parameters in $B(\xi_i)$ are fixed to the results from the PWA fit on the candidate events with $M_{\kpkm}$ in the range between 1.07 and 1.5~$\gevcc$.

In the PWA fit we include all the possible intermediate states listed in the PDG which can decay into  $\phi\eta$, $\phi\piz$, or $\piz\eta$ final states, and do not violate conservation of the quantum numbers $J^{PC}$.
To take into account the contribution from the coherent non-resonant component, the amplitudes modeled with the same sequential process and with a very broad and given $J^{PC}$ intermediate state are also included.
 The statistical significance of each individual process is obtained by examining the changes of the likelihood value and of the number of degrees of freedom between the cases with and without the corresponding process included in the fit, and only those processes with statistical significance larger than 5$\sigma$ are kept in the baseline solution.
Based on this solution, each dropped process is reintroduced individually and the corresponding statistical significance is estimated to verify that the process can be safely excluded.

Table~\ref{summa:Br} summarizes the dominant components in the baseline solution, their statistical significance, and the product branching fractions.
Beside the $\az-\fz$ mixing signal and the $\jpsi\to\phi\az$ electromagnetic decay, the dominant processes are those including intermediate states with masses at 1.68 and 1.9~$\gevcc$ on the $M_{\phi\eta}$ distribution.
There are additional contributions of $\rho(1570)$, $\rho(1900)$, $\rho(2050)$, and $\rho(2150)$ states on the $M_{\phi\piz}$ distribution, $\phi(1680)$ and $\phi(2170)$ states on the $M_{\phi\eta}$ distribution, $a_0(1450)$, $a_0(1950)$, $a_2(1320)$, and $a_2(1700)$ states on the $M_{\piz\eta}$ distribution~\cite{suppl}. These have small branching fractions and are strongly affected by the non-$\phi$ background; therefore, they are not reported in detail in this Letter.
The prominent structure at 1.4~$\gevcc$ on the $M_{\phi\piz}$ distribution can be well described by the non-$\phi$ background processes.
In the PWA, different $J^{PC}$ assignments for the intermediate states are investigated to determine their $J^{PC}$ individually.
A detailed comparison between the data and the MC projection indicates that the structure at 1.9~$\gevcc$ on the $M_{\phi\eta}$ distribution can not be described only by a new axial-vector state (named $h_1(1900)$ thereafter), but requires an additional vector state (named $X(2000)$ thereafter).
The values of mass and width of the intermediate states, summarized  in  Table~\ref{summa:Br}, are obtained by performing a scan of the likelihood values under different hypotheses and finding the minima of the distributions~\cite{suppl}. 
The comparison of the invariant mass and angular distributions between data and MC projections is shown in Fig.~\ref{PWAfiguure2125:2}.
The $\chi^{2}/nbin$ values for each distribution are indicated to estimate the goodness of the fits, where $nbin$ is the number of bins in each histogram and $\chi^{2}$ is defined as
 \begin{equation}
      \chi^{2} = \sum_{i=1}^{nbin}\frac{(n_i-\nu_i)^{2}}{n_i},
 \label{eq0}
 \end{equation}
 where $n_i$ and $\nu_i$ are the number of events for the data and the fit projection in the $i$-th bin, respectively.

     \begin{table}[htbp]
    \caption{The values of mass, width, product branching fraction ($\mathcal{B}$) and  statistical significance~($\sigma$) from the PWA fit for different subdecays, where the first uncertainties are statistical and the second systematic.}\label{summa:Br}
    \begin{center}
    \scriptsize
    \begin{tabular}{ l |c|c|c |c}
    \hline
    \hline
     Process &       M (MeV/$c^{2}$) &  $\Gamma$ (MeV)   & $\mathcal{B}$ ($10^{-6}$)  & $\sigma$ \\ \hline
     $\phi(1680)\piz$   &  $1668 \pm 7 \pm  25$          & $147 \pm 14 \pm 35$                    &  5.34 $\pm$ 0.25 $\pm$     0.88  & 36 \\
     $X(2000)\piz$   & $1996 \pm 11 \pm 30$ & $148 \pm 16 \pm 66$   &  2.76 $\pm$ 0.22  $\pm$  0.67 &17  \\
     $h_1(1900)\piz$      & $1911 \pm 6 \pm 14$ & $149 \pm 12 \pm 23$   &  9.29 $\pm$ 0.33 $\pm$  0.72  & 24\\
     $\phi a_{0}(980)_{\rm EM}$     &--                &--                    &  2.81 $\pm$ 0.24 $\pm$ 0.55   & 19 \\
     $\phi a_{0}(980)_{\rm mix}$    &--                &--                    &  2.73 $\pm$ 0.14 $\pm$ 0.19 & 32 \\
    \hline \hline
    \end{tabular}
    \end{center}
    \end{table}

Two categories of systematic uncertainties are considered in this analysis.
 The first category affects only the branching fraction measurements and includes uncertainties associated with the photon detection (1.0\% per photon)~\cite{BESIII:2010ank}, the MDC tracking (1.0\% per charged track)~\cite{BESIII:2011wmh}, the PID (1.0\% per kaon)~\cite{BESIII:2011ysp}, the kinematic fit (1.3\%), the $\phi$ mass resolution (0.9\%), the branching fractions of intermediate state decays (1.1\%)~\cite{ParticleDataGroup:2020ssz}, and the total number of $\jpsi$ events (0.4\%)~\cite{BESIII:2021cxx}.
 The uncertainty related to the 4C kinematic fit is estimated by correcting the helix parameters of the simulated charged tracks to compensate the resolution difference between data and MC simulation~\cite{BESIII:2012mpj}.
 The systematic uncertainties from the branching fractions of intermediate states in the subsequent decays are taken from the PDG~\cite{ParticleDataGroup:2020ssz}.
 The second category of uncertainties, which are mostly from the PWA fit procedure, affects both the branching fraction and the resonance parameters measurements.
To estimate these uncertainties, alternative fits with different scenarios are performed, and the resultant changes on the branching fractions and the resonance parameters are taken as the systematic uncertainties.
 Uncertainties from the BW parameterization are estimated by replacing the constant-width BW with the mass-dependent width.
Uncertainties associated with the resonance parameters, taken from the PDG and fixed in the fit, are estimated by individually varying the parameters by $\pm 1\sigma$  in the fit. The largest changes with respect to the nominal results are taken as the systematic uncertainties.
 To estimate the uncertainty due to the non-$\phi$ background model, alternative fits for the candidate events with  $M_{\kpkm}$ within [1.07, 1.3]~$\gevcc$ are carried out to model the background and the PWA fit for the signal candidate events.
The uncertainties associated with the additional resonances are estimated by alternative fits including the components $X(1750)$ on the $M_{\phi\eta}$ distribution and $\rho(1700)$ on the $M_{\phi\piz}$ distribution, which have the highest significances, but less than 5$\sigma$ with respect to the baseline solution.
 Assuming that all the sources of uncertainties are independent, the total uncertainties are estimated as the quadratic sum of the above individual values, shown in the supplemental material~\cite{suppl}.

  In summary, using 10 billion $\jpsi$ events collected with the BESIII detector operating at the BEPCII collider, a PWA of the $\jpsi \to \phi \piz \eta$ decay is performed for the first time using the covariant tensor amplitude method. Besides the well established vector strangeonium meson $\phi(1680)$, 
an  axial-vector meson $h_1(1900)$, with mass M = (1911 $\pm$ 6 $\pm$ 14)~$\mevcc$ and width $\Gamma$ = (149 $\pm$ 12 $\pm$ 23)~$\mev$, and a vector meson $X(2000)$, with mass M = (1996 $\pm$ 11 $\pm$ 30)~$\mevcc$ and width $\Gamma$ = (148 $\pm$ 16 $\pm$ 66) $\mev$, are observed for the first time, with statistical significances of 24$\sigma$ and 17$\sigma$, respectively. 
The $h_1(1900)$ meson can be assigned as the $h_1(2P)$ strangeonium state, since its mass and width agree well with the theoretical predictions~\cite{Barnes:2002mu,Li:2020xzs}. The $X(2000)$ state can be associated to the $\phi(3S)$ or $\phi(2D)$ states, but its mass and width have large discrepancies with the theoretical predictions~\cite{Barnes:2002mu,Li:2020xzs}.
The observation of the new $h_1(1900)$ and $X(2000)$ states provides important input to establish the strangeonium $s\bar{s}$ spectrum, and calls for further theory work.

The decay branching fractions of $\jpsi$ $\to$ $\phi \fz$ $\to \phi \az$ and the corresponding EM process $\jpsi$ $\to$ $\phi \az$ are
measured with substantially improved precision. The mixing intensity of $\fz-\az$ ($\xi_{fa}$) is calculated to be ($0.85 \pm 0.04 \pm 0.25$)\%, which is consistent with the destructive solution in the previous measurement~\cite{BESIII:2018ozj}. It can be used to estimate the coupling constants of the $\fz$ and $\az$ resonances~\cite{Achasov:2016wll}. 
Moreover, the two-fold ambiguity from the previous analysis~\cite{BESIII:2018ozj} is resolved here by performing the PWA of $\jpsi \to \phi \piz \eta$. This provides crucial constraints for models describing the internal structure of $\az$ and $\fz$.

The BESIII Collaboration thanks the staff of BEPCII, the IHEP computing center and the supercomputing center of USTC for their strong support. This work is supported in part by National Key R\&D Program of China under Contracts Nos. 2020YFA0406400, 2020YFA0406300; National Natural Science Foundation of China (NSFC) under Contracts Nos. 11635010, 11735014, 11835012, 11935015, 11935016, 11935018, 11961141012, 12022510, 12025502, 12035009, 12035013, 12192260, 12192261, 12192262, 12192263, 12192264, 12192265, 11335008, 11625523, 12035013, 11705192, 11950410506, 12061131003, 12105276, 12122509, 12235017, 12205255; the Chinese Academy of Sciences (CAS) Large-Scale Scientific Facility Program; the CAS Center for Excellence in Particle Physics (CCEPP); Joint Large-Scale Scientific Facility Funds of the NSFC and CAS under Contract No. U1832207, U1732263, U1832103, U2032111, U2032105; CAS Key Research Program of Frontier Sciences under Contracts Nos. QYZDJ-SSW-SLH003, QYZDJ-SSW-SLH040; 100 Talents Program of CAS; The Institute of Nuclear and Particle Physics (INPAC) and Shanghai Key Laboratory for Particle Physics and Cosmology; European Union's Horizon 2020 research and innovation programme under Marie Sklodowska-Curie grant agreement under Contract No. 894790; German Research Foundation DFG under Contracts Nos. 455635585, Collaborative Research Center CRC 1044, FOR5327, GRK 2149; Istituto Nazionale di Fisica Nucleare, Italy; Ministry of Development of Turkey under Contract No. DPT2006K-120470; National Research Foundation of Korea under Contract No. NRF-2022R1A2C1092335; National Science and Technology fund of Mongolia; National Science Research and Innovation Fund (NSRF) via the Program Management Unit for Human Resources \& Institutional Development, Research and Innovation of Thailand under Contract No. B16F640076; Polish National Science Centre under Contract No.2019/35/O/ST2/02907; The Swedish Research Council; U. S. Department of Energy under Contract No. DE-FG02-05ER41374.


\end{document}